\documentclass[trackchanges]{aastex701}

\usepackage{graphicx}
\usepackage[version=4]{mhchem}

\begin{document}

\title{Small-Body Science with the Nautilus Space Observatory: \\
From Cislunar Space Resilience to Mapping the
Kuiper-Belt-to-Oort-Cloud Transition}

\author[0000-0003-2415-2191]{J. de Wit}
\affiliation{Massachusetts Institute of Technology, Cambridge, MA, USA}
\email[show]{jdewit@mit.edu}

\author[0000-0001-9892-2406]{A. Y. Burdanov}
\affiliation{Massachusetts Institute of Technology, Cambridge, MA, USA}
\email{burdanov@mit.edu}

\author[0000-0002-1052-6749]{P. McGill}
\affiliation{Space Science Institute, Lawrence Livermore National Laboratory, 7000 East Avenue, Livermore, CA 94550, USA}
\email{mcgill5@llnl.gov}

\author[orcid=0000-0003-3714-5855]{D. Apai}
\affiliation{Steward Observatory, The University of Arizona, 933 N. Cherry Avenue, Tucson, AZ 85721, USA}
\affiliation{Lunar and Planetary Laboratory, University of Arizona, 1629 E. University Boulevard, Tucson, AZ 85721, USA}
\affiliation{Alien Earths Team, NASA ICAR/NExSS, USA}
\email{apai@arizona.edu}

\author[orcid=0000-0002-4894-193X]{S. N.~Hasler}
\affiliation{Space Telescope Science Institute, 3700 San Martin Drive, Baltimore, MD 21218, USA}\affiliation{Massachusetts Institute of Technology, Cambridge, MA, USA}\email{shasler@stsci.edu}

\author[orcid=0000-0001-6294-4523]{S. Cambioni}
\affiliation{Massachusetts Institute of Technology, Cambridge, MA, USA}\email{cambioni@mit.edu}

\begin{abstract}

A new generation of space-based observatories can transform small-body science from the local Earth–Moon environment to the Solar System’s dynamical frontier, where planetary control gives way to Galactic tides and stellar encounters. Mapping this frontier would reveal how planetesimals were scattered during giant-planet formation, how many primordial bodies survived in distant reservoirs, and whether large objects remain undiscovered in the inner Oort Cloud. At the same time, meter- to decameter-scale natural objects traversing the Earth–Moon system are frequent but poorly characterized. As human and robotic activity expands, their operational relevance depends not simply on abundance but on object size, encounter geometry, exposed infrastructure, warning time, and available response. Lunar impacts and impact-generated ejecta provide one class of events, while fading objects with unresolved Earth or lunar impact probabilities provide another. Here, we explore a two-layer small-body program combining a 4.0 m-diameter Nautilus Quattro Imaging (NQI) unit with an illustrative distributed cislunar layer of $\sim$50 cm-class wide-field optical telescopes. The cislunar layer would characterize the nearby natural-object population, support rapid classification and selective custody of high-interest objects, and monitor lunar impacts and ejecta; the NQI deep-search layer would provide precise astrometry and synthetic tracking for faint-object recovery and distant-Solar-System surveys. Together, the two layers would constrain the small-impactor population, resolve fading planetary-defense targets, and survey the distant Solar System to \(V\sim30\), enabling tests of the Kuiper-Belt-to-Oort-Cloud transition while advancing Earth–Moon space resilience.

\end{abstract}

\vspace{-1.5cm}

%=========================

\vspace*{-1cm}\section*{The Nautilus Space Observatory}

The Nautilus Space Observatory is a proposed scalable space-observatory architecture designed to combine large collecting area, modular deployment, and flexible survey strategies across a broad range of astrophysical and planetary-science applications \citep{2019BAAS...51g.141A}. This white paper is part of the Nautilus Science Case Workshop series \citep[see, e.g.,][]{2026arXiv260700268W,2026arXiv260630706F,2026arXiv260628523W,2026arXiv260626462L,2026arXiv260626214P} and explores how such an architecture could enable faint moving-object discovery, tracking, characterization, and population inference across the Solar System. 
We focus on a staged small-body program linking cislunar impactor and lunar-ejecta science, follow-up-limited planetary defense, and deep surveys of the Kuiper-Belt-to-Oort-Cloud transition.

\section{Scientific Context and Problem Statement}

\begin{figure*}
\hspace*{-8mm}
\includegraphics[width=1.1\textwidth]{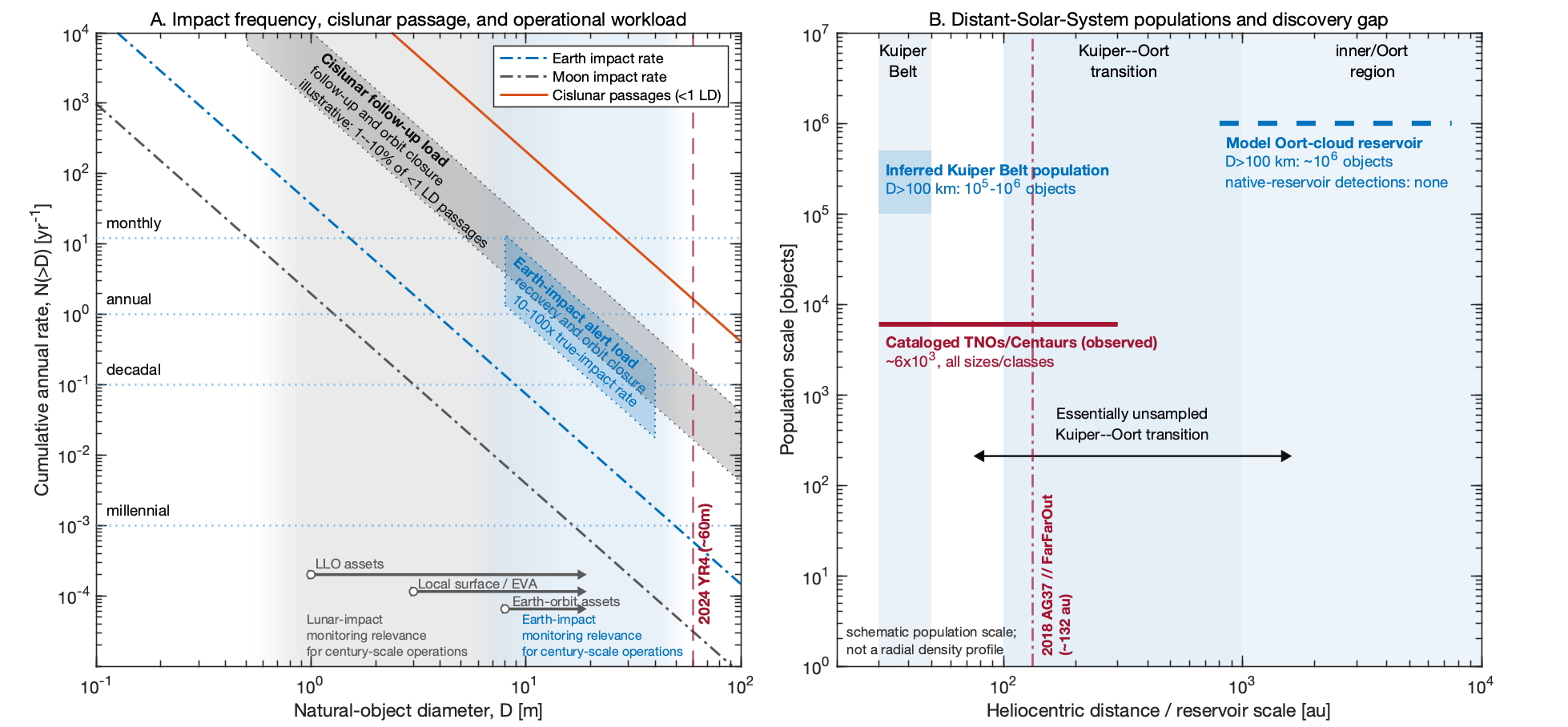}
\caption{
\textbf{Population and operational motivation for a two-layer Nautilus small-body program.}
\textbf{Left (A):} Cumulative annual rates of natural objects as a function of diameter. The Earth-impact curve follows \citet{Brown2002}; the lunar-impact curve applies an Earth--Moon collision-rate scaling, while the upper curve gives the estimated physical rate of passages within one lunar distance (LD) of Earth. The gray band illustrates a possible cislunar follow-up and orbit-closure workload if approximately 1--10\% of physical passages require continued processing; this fraction is a sensitivity range rather than an empirical workload measurement. The blue band shows the distinct alert-closure workload associated with fading potential Earth impactors whose nonzero impact probabilities require deeper recovery and orbit refinement. As demonstrated by 2024 YR4, such objects can become inaccessible from the ground while retaining nonzero impact probabilities \citep{deWit2026}. Illustratively, if only a few true decameter-scale Earth impactors occur during the twenty-first century, a follow-up load $\sim10$--$100\times$ the true-impact rate would imply deep recovery and orbit closure for tens to hundreds of ultimately non-impacting candidates---an operational cadence of up to several targets per year rather than one per decade. The multiplier is a sensitivity range rather than an empirical workload measurement. Horizontal asset markers summarize preliminary size ranges over which lunar-impact monitoring may become relevant to LLO, local surface/EVA, and Earth-orbit operations. These are illustrative sensitivities, not hard hazard thresholds, and their vertical positions carry no rate information. Shading emphasizes regimes combining potential operational relevance with event frequency over a century-scale planning horizon. Actual relevance depends on impact geometry, ejecta properties, asset vulnerability, object-characterization uncertainty, and survey/follow-up performance. The position of 2024 YR4 ($\sim60$~m) is shown for scale.
\textbf{Right (B):} Schematic comparison of observed and estimated distant-Solar-System populations. The cataloged population of approximately $6\times10^3$ TNOs and Centaurs is shown separately from the inferred Kuiper Belt population and modeled Oort-cloud reservoir with order $10^6$ objects with $D>100$\,km, despite the absence of any direct native-reservoir detection~\citep{Morbidelli2020}. The discovery distance of 2018~AG37 (FarFarOut), approximately 132\,au, is marked for reference; the region extending from roughly $10^2$ to $10^3$\,au remains essentially unsampled and may contain the transition from a planet-shaped trans-Neptunian disk to a more isotropic reservoir~\citep{gladman2021ARA&A..59..203G}. 
}
\label{fig:population_motivation}
\end{figure*}

Asteroid and small-body science is entering a new observational regime as next-generation surveys and space-based observatories extend our ability to discover, track, and characterize faint moving objects from the Earth--Moon environment to the distant Solar System. This creates an opportunity to connect two regimes usually treated separately: frequent meter- to decameter-scale natural objects traversing cislunar space, and faint distant populations that preserve the fossil record of Solar System formation. These observational gaps are summarized in Figure~\ref{fig:population_motivation}.

The local population is becoming increasingly important as human and robotic activity expands throughout the Earth–Moon system. Small natural objects are far more frequent than globally hazardous asteroids~\citep{Brown2002,Brown2013}. Some strike the Moon, while others pass through cislunar space on trajectories relevant to planetary defense or emerging infrastructure. On Earth, the atmosphere filters much of this flux; on the Moon, the lack of atmosphere and low gravity allow impacts to generate ejecta reaching surface, orbital, and—under favorable conditions—broader cislunar environments ~\citep{Wiegert2025,He2026}. The operational relevance is strongly scale- and asset-dependent: smaller, more frequent events may matter locally or in low lunar orbit, whereas progressively larger and rarer events are required to affect broader infrastructure. In this paper, \emph{cislunar space resilience} refers specifically to this natural-object dimension of the broader Earth–Moon operating environment.

Figure~\ref{fig:population_motivation} places these regimes in a common frequency–size framework. The Earth-impact curve follows the cumulative terrestrial-impact relation of \citet{Brown2002}, with the lunar rate obtained from the corresponding Earth–Moon collision-rate scaling; physical passages within one lunar distance are orders of magnitude more frequent than impacts. Meter-scale lunar impacts occur on roughly annual or shorter timescales, while decameter impacts are far rarer and more uncertain. The much larger passage population creates a distinct observing problem: most objects are benign, so the relevant task is rapid detection and classification followed by selective continued observation or recovery when trajectories or consequences remain unresolved. Bright Earth/Moon backgrounds, rapidly changing illumination, large apparent motions and parallax, and Earth–Moon dynamics make that task particularly demanding. Complementary space-based viewpoints could improve visibility, parallax, orbit refinement, and continuity, motivating the distributed Layer-1 architecture explored below.

Recent observations of asteroid 2024 YR4 illustrate the complementary follow-up problem~\citep{Rivkin2026,deWit2026}. Wide-field facilities such as Rubin, NEO Surveyor, Roman, NEOMIR-like concepts, and other surveys will substantially expand small-object discovery~\citep{Ivezic2019,Mainzer2023,Conversi2024,Holler2025}, but some high-interest objects will remain limited by short discovery arcs, unfavorable geometry, or fading beyond ground-based reach. JWST reached \(V\sim30.5\) for 2024 YR4, extended its observational arc by eight months, and resolved the remaining 2032 lunar-impact probability roughly two years before ground-based recovery would have been possible. Although \citet{Burdanov2025} highlighted JWST’s potential as a planetary-defense asset—a capability later demonstrated for 2024 YR4---the observatory was not designed for routine rapid-response moving-object follow-up. Nautilus could complement discovery surveys through selective cislunar tracking and deep faint-object recovery, accelerating orbit closure for the most consequential unresolved cases.

Beyond hazard resolution, any monitored object that ultimately struck the Moon would provide a uniquely calibrated impact experiment \citep{2010Sci...330..463C}. Its pre-impact orbit and known impact time, location, energy, and geometry, combined with observations of the flash, ejecta, and seismic response, could test impact-scaling laws, constrain which lunar terrains can launch material onto Earth-crossing trajectories, calibrate future lunar seismic networks and the interpretation of impact-generated seismic events, and improve assessments for safe lunar surface and orbital operations.

The same observational capabilities are also required at the opposite end of the Solar System. The transition from the Kuiper Belt to the Oort Cloud remains the last largely unexplored region of the Solar System. Trans-Neptunian objects (TNOs) and more distant small bodies preserve a record of planetesimal formation, giant-planet migration, Galactic perturbations, stellar encounters, and the early dynamical environment of the Sun~\citep{duncan1997Sci...276.1670D,Morbidelli2020,gladman2021ARA&A..59..203G}. Yet the known distant small-body population is strongly biased toward objects near the ecliptic and bright enough to be detected in individual exposures, leaving distant, small, low-albedo, and high-inclination bodies poorly sampled~\citep{gladman2021ARA&A..59..203G}. A transition from an ecliptic-dense population to a more isotropic reservoir should occur where planetary perturbations give way to Galactic tides and stellar encounters; detecting it would locate the boundary of the planet-shaped Solar System.

The missing capability is a deep, repeated, wide-field survey combining shifted-and-stacked apparent-magnitude sensitivity reaching $V\sim30$, ecliptic and off-ecliptic coverage, and multi-month astrometric baselines sufficient to detect, recover, and dynamically classify objects at hundreds to thousands of astronomical units (au). This transition has not yet been observed. Although long-period comets provide indirect evidence for the Oort Cloud, no object has yet been detected in its native distant reservoir. Theoretical models nevertheless predict a vast population that may include $\sim10^6$ bodies with diameters $D\gtrsim100$ km and, potentially, Charon- to Pluto-scale or larger objects~\citep{Stern1991,Morbidelli2020,gladman2021ARA&A..59..203G}. The inner Oort Cloud is expected to begin at hundreds to thousands of au, while most known TNOs lie within $\sim100$ au~\citep{gladman2021ARA&A..59..203G}. The question is therefore not only whether rare large bodies exist at hundreds to thousands of au, but whether the broader distant population reveals a measurable dynamical transition from the Kuiper Belt to the Oort Cloud.

%=========================

\section{Science Objectives}

We propose a scalable two-layer small-body program built around the Nautilus Space Observatory. The cislunar component would use modest wide-field optical telescopes in complementary high-Earth and cislunar viewing geometries for rapid detection and classification, selective continued tracking, and lunar-impact monitoring, while a 4.0\,m-diameter Nautilus Quattro Imaging (NQI) unit would provide the depth required for faint planetary-defense recovery and distant-Solar-System surveys. The components are complementary: distributed wide-field coverage, rapid cadence, and separated viewing geometries provide local discovery, parallax, and follow-up continuity, whereas the NQI unit supplies repeated deep imaging, precise astrometry, and motion-aware processing for targeted recovery and Kuiper-Belt-to-Oort-Cloud population studies. The program is organized around three core science objectives:

\begin{enumerate}

\item \textbf{Characterize natural objects and impact events in the Earth--Moon system.}
Measure the flux, size distribution, orbital properties, and temporal variability of meter- to decameter-scale objects traversing cislunar space. For objects that ultimately strike the Moon, connect the known pre-impact trajectory to the impact, ejecta, and seismic response. As a demanding Layer-1 design case, we consider detection and tracklet formation for $\sim1$\,m potential lunar impactors with $\sim$6--12\,hr warning, a sensitivity relevant to local lunar and low-lunar-orbit operations rather than a universal hazard threshold.

\item \textbf{Resolve ambiguous impact probabilities for faint planetary-defense targets.}
Recover, prioritize, and refine the orbits of small objects that fade beyond ground-based limits before their Earth or lunar impact probabilities can be driven toward practical 0-or-1 closure. Because residual probabilities can remain non-negligible even as targets fade, efficient triage and deep recovery of a larger population of ultimately non-impacting candidates become an important complement to identifying the rare true impactor.

\item \textbf{Map the Kuiper-Belt-to-Oort-Cloud transition.}
Measure the orbital, inclination, color, and size distributions of faint distant Solar System objects as a function of distance, test for a transition from a planet-shaped disk toward a more isotropic reservoir, and constrain the abundance of large inner-Oort-Cloud bodies.

\end{enumerate}

Together, the first two objectives probe the nearby small-body population through complementary modes: wide-field characterization, rapid classification, selective continued tracking, lunar-impact science, and deep orbit closure for exceptional fading targets. The third applies the same moving-object capabilities to a fundamental Solar System origins problem. The program would deliver public catalogs of cislunar natural objects, lunar-impact events, faint planetary-defense follow-up measurements, and distant Solar System discoveries, together with quantified detection and recovery selection functions, astrometry, photometry, and orbit solutions required for population inference. The same imaging would provide ancillary time-domain searches for active and interstellar small bodies, stellar variability, microlensing, compact binaries, and other transient phenomena.

%=========================
\section{Data Requirements}

\begin{figure*}
\hspace*{-8mm}
\includegraphics[width=1.1\textwidth]{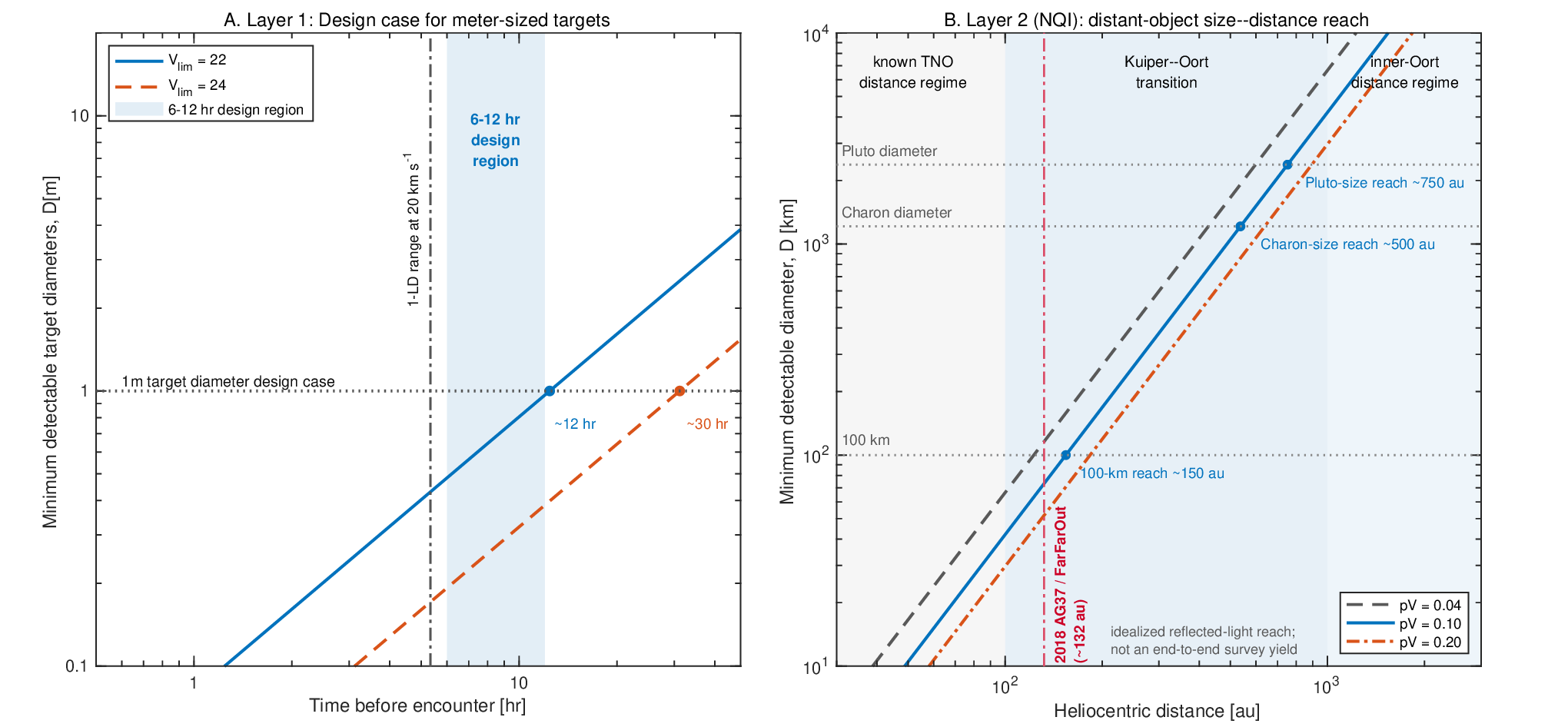}
\caption{
\textbf{First-order sensitivity reach and representative design cases for the two observing layers.}
\textbf{Left (A):} Minimum detectable diameter for an approaching cislunar object as a function of time before encounter, shown for motion-compensated limiting apparent magnitudes of $V_{\rm lim}=22$ and 24. The calculation assumes a geometric albedo of $p_V=0.10$, heliocentric distance $r=1$\,au, opposition geometry, and a representative closing speed of 20\,km\,s$^{-1}$ to convert observer--object range into warning time. The horizontal line and shaded region mark an illustrative 1\,m / 6--12\,hr Layer-1 design case motivated by local lunar and low-lunar-orbit applications rather than a universal operational threshold. Under these idealized assumptions, a 1\,m object would be detectable approximately 12\,hr before encounter at $V_{\rm lim}=22$ and approximately 30\,hr before encounter at $V_{\rm lim}=24$; the vertical reference line marks the time corresponding to a range of one lunar distance at the adopted closing speed.
\textbf{Right (B):} Minimum detectable diameter in the distant Solar System for a shifted-and-stacked apparent-magnitude depth of $V=30$ and representative geometric albedos of $p_V=0.04$, 0.10, and 0.20, assuming the observer--object distance is approximately equal to the heliocentric distance. Shaded regions indicate the main sampled trans-Neptunian regime, the Kuiper--Oort transition, and the inner-Oort distance regime. The vertical line marks 2018~AG37 (FarFarOut) at approximately 132\,au, while horizontal guides mark diameters of 100\,km, Charon, and Pluto. For the representative $p_V=0.10$ case, the corresponding idealized reaches are approximately 150\,au for a 100\,km object, 500\,au for a Charon-size object, and 750\,au for a Pluto-size object. Both panels show idealized reflected-light sensitivity; they are photometric-reach estimates rather than end-to-end survey-yield predictions.
}
\label{fig:capability_reach}
\end{figure*}

Figure~\ref{fig:capability_reach} translates representative science and design cases into first-order sensitivity benchmarks for each layer. Their distinct requirements for observing geometry, exposure strategy, cadence, and processing are described below. The numerical values are preliminary performance drivers rather than optimized mission requirements and should be refined through end-to-end simulations.

\subsection{Layer 1: Cislunar monitoring and selective tracking}

Layer~1 would characterize meter- to decameter-scale natural objects traversing the Earth--Moon system through rapid detection and classification, selective continued tracking, and monitoring of lunar impacts and their ejecta. As a demanding design case, we consider sensitivity to favorable-phase $\sim1$\,m potential lunar impactors with hours-scale warning. Such a capability could support pre-impact trajectory refinement and estimates of impact time, location, and energy for selected events, supporting targeted scientific observations and, where relevant, operational response. The same observing system would necessarily detect a much larger population of non-impacting or still-ambiguous objects, including possible Earth impactors and bodies passing through cislunar space.

An illustrative implementation would use approximately five to eight $\sim50$\,cm-class wide-field telescopes distributed across complementary high-Earth and/or cislunar viewing geometries. These values should be regarded as an initial architecture point rather than a derived requirement: the final number of nodes, aperture, field of view, and orbital configuration depend jointly on sensitivity, sky coverage, cadence, viewing geometry, and the fraction of detected objects requiring continued follow-up. The concept builds on the wide-field imaging and lunar-resonant high-Earth-orbit heritage of TESS \citep{Ricker2015}, while extending it toward larger apertures, distributed viewpoints, and motion-aware processing for rapidly moving natural objects.

At representative cislunar encounter velocities, nearby meter-scale objects can move by several to more than ten arcseconds per second. Elementary exposures must therefore be short enough to control trailing, plausibly $\sim0.2$--2\,s depending on apparent rate, pixel scale, and image quality. Sensitivity would be accumulated through motion-compensated sequences totaling approximately 30--120\,s, placing corresponding requirements on low-noise, high-cadence focal-plane readout and onboard processing. In a broad optical discovery band spanning approximately 0.4--1.0\,$\mu$m, a useful first-order target is a motion-compensated, $V$-equivalent limiting magnitude of $V_{\rm lim}\sim21$--22 in broad-survey mode, with targeted or longer integrations approaching $V_{\rm lim}\sim23$--24. As illustrated in Figure~2, these sensitivities place a favorable-phase, $p_V=0.10$, 1\,m object in the hours-warning regime and therefore provide a useful stress case for Layer-1 instrument design.

Separated viewpoints would provide complementary Sun/Earth/Moon visibility and parallax constraints on object distance and velocity while reducing common viewing-geometry losses. Observations over minutes to hours would form and associate tracklets, with longer baselines used selectively to refine orbits and maintain knowledge of high-interest objects. Onboard processing would identify candidates and return prioritized astrometry, photometry, trajectory uncertainties, and image cutouts while retaining calibrated elementary exposures for selective downlink. The final architecture must balance sensitivity, field of view, apparent-rate coverage, revisit cadence, geometric completeness, communications, launch opportunities, station keeping, and the radiation and space-weather environment.

\newpage

\subsection{Layer 2: Deep search and characterization}

The NQI unit would provide two complementary capabilities: discovery of faint distant Solar System objects and targeted recovery of faint planetary-defense objects. A broad, high-throughput optical discovery band is required, together with at least one additional color band; three bands are preferred for population-level color classification, as the orbits and colors of these objects hold clues to the history of the Solar System \citep{gladman2021ARA&A..59..203G}.

For planetary-defense follow-up, the key requirement is deep imaging of predicted target locations and uncertainty regions to recover objects that have faded beyond ground-based limits and deliver astrometry rapidly enough to update impact probabilities. This mode does not require continuous survey coverage, but it does require flexible scheduling, accurate pointing, repeated exposures suitable for synthetic tracking, and rapid orbit-reporting products. It would operate primarily as a target-of-opportunity mode using a small fraction of an NQI unit's observing time. The illustrative alert-closure workload in Figure~\ref{fig:population_motivation} motivates designing for multiple recovery campaigns per year while retaining sufficient flexibility for rarer high-priority cases; the actual cadence will depend on future discovery rates, self-recovery by survey facilities, and follow-up performance.

For the distant Solar System survey, the target depth is shifted-and-stacked sensitivity reaching $V\sim30$. For objects at hundreds to thousands of astronomical units, parallactic motion can dominate intrinsic orbital motion, making accurate timing, astrometric calibration, observatory ephemerides, and parallax-aware trajectory modeling essential. Each discovery epoch would consist of repeated short exposures on a given field over several hours or a substantial fraction of a day, enabling digital tracking over trial trajectories while accumulating the required depth. Short-baseline revisits over days to weeks would confirm candidates and establish initial motion solutions. Revisits after $\sim4$ months would exploit the changed parallax vector to constrain heliocentric distance and support cross-epoch linking, while additional observations over six months or longer would refine orbital elements and dynamical classifications.

A central Layer-2 design question is therefore what minimum combination of depth, field of view, survey area, ecliptic and off-ecliptic coverage, and revisit cadence is required to distinguish a flattened trans-Neptunian population from one evolving toward isotropy with increasing heliocentric distance, while retaining sensitivity to rare large bodies at hundreds to thousands of au. The required footprint, sample size, and cadence should be established through end-to-end population, synthetic-tracking, and recovery simulations.

\subsection{Minimum data products}

The minimum required data products are:

\begin{itemize}

\item calibrated individual exposures with accurate time stamps, pointing and distortion solutions, spacecraft ephemerides, background and uncertainty estimates, and quality flags;

\item moving-object products including detections, apparent motions, tracklets, cross-epoch or cross-node associations, astrometry, photometry, trajectory uncertainties, and preliminary orbit solutions or classifications;

\item rapid-alert and public legacy products, including prioritized follow-up information for high-interest cislunar and planetary-defense objects, final catalogs and orbit classifications, colors or limits, and selection functions derived from injection--recovery and completeness analyses.

\end{itemize}

%=========================
\section{Analysis and Interpretation}

The analysis will convert repeated imaging into orbit-linked catalogs of cislunar natural objects, lunar-impact events, planetary-defense targets, and distant Solar System populations, with quantified completeness and uncertainties.

\subsection{Cislunar detection, classification, and impact interpretation}

Cislunar detections will be linked across exposures and separated viewing geometries and fit with dynamical models appropriate for the Earth--Moon system. The analysis must account for spatially variable Earth/Moon backgrounds, changing illumination and phase geometry, large apparent motions and parallax, and nontrivial three-body dynamics. Cross-node associations will constrain distance and velocity, reject false positives, and help distinguish lunar impactors, possible Earth impactors, temporarily captured objects, and non-impacting flybys.

The resulting catalog will include astrometry, photometry, orbit solutions and uncertainties, trajectory classifications, impact probabilities where relevant, and quantified detection, classification, and follow-up performance. For lunar-impacting objects, performance will be characterized as a function of size, approach direction, warning time, phase angle, background, and observing geometry. Injection--recovery and orbit-determination simulations will quantify which parts of this parameter space can be classified rapidly and which require selective tracking or follow-up.

For objects observed prior to lunar impact, photometry and orbit-derived velocity would be combined with observations of the impact flash, ejecta, and seismic response to constrain impactor size and mass, impact energy, and ejecta production. These estimates would remain model-dependent because visible photometry alone does not uniquely determine diameter and albedo, while the high-speed fragment distribution remains an important uncertainty in translating ejecta mass into operational exposure. At the population level, these measurements would constrain the flux, size distribution, orbital distribution, and temporal variability of meter- to decameter-scale natural objects in the Earth--Moon system, while individual calibrated impacts would test models of ejecta environments relevant to lunar-surface, low-lunar-orbit, and, for sufficiently energetic and favorably configured events, broader cislunar operations.

\subsection{Faint-object follow-up and orbit refinement}

The primary output is astrometry delivered rapidly enough to update impact probabilities and guide subsequent observations. As demonstrated by the 2024 YR4 use case, the objective is the recovery and orbit refinement of specific objects that have become inaccessible from the ground, with the NQI layer extending toward $V\sim30$ when required.

\subsection{Distant Solar System population inference}

Distant-object candidates will be identified through digital tracking and linked across the multi-epoch observing sequence described in Section~3.2. Parallax-aware trajectory fitting will constrain distances and orbital elements, allowing classification as Centaurs, TNOs, detached or scattered-disk objects, or candidate inner-Oort-Cloud bodies. Photometry and broad-band colors will be measured along the motion solutions and compared across dynamical classes.

Injection--recovery tests will quantify completeness as a function of magnitude, motion, color, background, sky position, and cadence. These selection functions will be used to infer the underlying inclination, orbit, color, size, and number-density distributions and to test whether the distant population evolves from a flattened trans-Neptunian disk toward a more isotropic Oort-Cloud-like reservoir with increasing heliocentric distance.

\subsection{Success criteria}

A successful analysis would deliver:

\begin{itemize}

\item a completeness-characterized catalog of cislunar natural objects, including tracklets, orbit solutions, trajectory classifications, and follow-up performance as a function of object size, approach geometry, and warning time;

\item empirical constraints on the flux, size, and orbital distributions of cislunar natural objects, together with calibrated impact-energy, ejecta, and environmental constraints for observed lunar-impact events;

\item rapid-recovery astrometry and orbit refinement for faint planetary-defense targets sufficient to reduce or resolve ambiguous impact probabilities;

\item a completeness-characterized catalog of distant Solar System objects to $V\sim30$, enabling tests of the Kuiper-Belt-to-Oort-Cloud transition and constraints on large inner-Oort-Cloud bodies, incl. meaningful upper limits.

\end{itemize}

%=====================================================
\section{Relevant Science Requirements}

Table~\ref{tab:scireq} consolidates preliminary science and performance drivers for the two-layer Nautilus small-body program. In both layers, the final architecture should be established through end-to-end simulations.

%=========================

\section{Relevance to Nautilus and Mission Class}

The concept is relevant to Nautilus because a NQI layer provides the deep, stable, and flexibly scheduled imaging required for distant-Solar-System surveys and faint-object recovery, while its scalable mission framework can accommodate coordination with a distributed cislunar monitoring layer.

\textbf{Cislunar layer:} A candidate homogeneous implementation would place several wide-field optical telescopes in complementary high-Earth and/or cislunar geometries. Lunar-resonant high-Earth orbits are particularly attractive starting points because of the operational heritage of TESS and their favorable combination of sky access, stability, and Earth--Moon viewing geometry \citep{Gangestad2013,Dichmann2013,Ricker2015}. Multiple viewpoints could improve directional coverage, reduce common Sun/Earth/Moon visibility losses, and provide parallax for rapid range and orbit estimation. Heterogeneous orbit families may provide additional geometric leverage. The illustrative $\sim5$--8 node, $\sim50$\,cm-class configuration considered here should therefore be regarded as a design point rather than a requirement; astrodynamical, photometric-yield, linking, and orbit-determination simulations should establish the minimum architecture needed to achieve the desired science performance.

\textbf{Deep-search layer:} A single NQI unit could support targeted planetary-defense recovery and a focused distant-Solar-System survey, with planetary-defense observations operating primarily as flexible target-of-opportunity campaigns between broader survey and astrophysics programs. Additional NQI units would increase survey speed, cadence, sky coverage, and follow-up flexibility, but are not required to demonstrate the central science case.

\textbf{Relevant Class:} The concept is naturally staged. A Probe-class implementation could demonstrate either the cislunar moving-object layer or a single NQI deep-search pathfinder, allowing the principal sensitivity, cadence, linking, and orbit-recovery assumptions to be tested before scaling. A Flaglet-class implementation could combine both components into a coordinated program, coupling rapid wide-field characterization of the Earth--Moon environment with deep recovery and a statistically meaningful distant-Solar-System survey. A larger distributed architecture would improve geometric coverage, resilience, and survey speed if warranted by end-to-end performance simulations, but is not assumed as a prerequisite for the core science.

%=========================

\section{Relevance to NASA and Astrophysics Strategy}

This science case intersects NASA priorities in small-body science, planetary defense, lunar exploration, Solar System origins, and the operation of an increasingly active Earth--Moon environment. It is complementary to existing planetary-defense, meteoroid-environment, lunar-impact, and space-domain-awareness capabilities. 

As NASA moves toward recurring lunar missions and sustained surface and orbital activity~\citep{NASA2025MoonToMars,NASA2026MoonBase}, the \textit{Moon to Mars Architecture} identifies the natural environment and advance knowledge of operational hazards as relevant considerations for lunar exploration. The proposed cislunar layer would address one component of this broader problem by measuring and classifying natural objects traversing the Earth--Moon system and, for selected lunar-impact events, connecting pre-impact trajectories to impact location, energy, ejecta, and subsequent environmental response. Lunar-impact relevance is strongly scale-, geometry-, and asset-dependent: smaller events may be most consequential for nearby surface or low-lunar-orbit operations, whereas effects reaching broader cislunar or Earth-orbit infrastructure require progressively larger and more favorable events. Measurements of actual impacts would provide both small-body science and empirical calibration of transient lunar-ejecta models.

The planetary-defense relevance is similarly direct. \textit{Origins, Worlds, and Life}, the 2023--2032 Planetary Science and Astrobiology Decadal Survey, emphasizes improved detection, tracking, and characterization of the largely unknown sub-140\,m NEO population \citep{NASEM2023}. Rubin, NEO Surveyor, and other discovery facilities will substantially enlarge the known population, but some high-interest objects may fade beyond routine ground-based reach before their impact probabilities are fully resolved. Asteroid 2024 YR4 provides a direct example: deep JWST imaging extended its observational arc and resolved the remaining lunar-impact probability well before ground-based recovery would otherwise have been possible \citep{deWit2026, 2026JAnSc..73....8M}. An NQI unit reaching approximately $V\sim30$, with flexible scheduling and motion-aware processing, could provide a dedicated complement to discovery surveys for such exceptional faint-target recovery and orbit-closure cases.

The distant-Solar-System program addresses a separate but equally strong planetary-science objective. Measurements extending beyond the trans-Neptunian region would constrain the dynamical structure, physical properties, and orbital subpopulations of distant small bodies \citep{NASEM2023}. A deep, repeated survey spanning ecliptic and off-ecliptic fields could test models of planetesimal formation, giant-planet migration, scattering, Galactic perturbations, and stellar encounters; search for the transition from a planet-shaped trans-Neptunian disk toward a more isotropic reservoir; and constrain the abundance of large inner-Oort-Cloud bodies.

The concept is also relevant to NASA Astrophysics and ASTRA because its rapid-cadence and deep-imaging modes retain broad astrophysical utility, from interstellar and active small bodies to stellar variability, microlensing, compact binaries, and other transients. Planetary-defense recovery could be interleaved with deep surveys and general-astrophysics programs, while the scalable architecture would exercise autonomous moving-object detection, onboard processing, and coordinated observations.

%% Please use the acknowledgment and contribution environments. This will 
%% be anonomyized when the "anonymous" style option is used. 
\begin{acknowledgments}
The authors thank the Heising-Simons Foundation for supporting the Nautilus Science Case Workshop. JdW thanks Richard Binzel, Deborah Woods, and Herbert Viggh for their valuable input. PM acknowledges that this work was performed in part under the auspices of the U.S. Department of Energy by Lawrence Livermore National Laboratory under the contract DE-AC52-07NA27344, LLNL-CONF-2020817.
\vspace{-5mm}
\end{acknowledgments}


%\appendix

\bibliography{sample701}{}
\bibliographystyle{aasjournalv7}

% Compact fixed-width table compatible with AASTeX 7.0.1 / TeX Live 2025.
% Keep the table body free of blank lines between \startdata and \enddata.
\providecommand{\reqname}[1]{\parbox[t]{0.14\textwidth}{\raggedright\strut #1\strut}}
\providecommand{\reqLone}[1]{\parbox[t]{0.38\textwidth}{\raggedright\strut #1\strut}}
\providecommand{\reqLtwo}[1]{\parbox[t]{0.38\textwidth}{\raggedright\strut #1\strut}}

\begin{deluxetable*}{lll}
\tabletypesize{\scriptsize}
\tablecaption{Preliminary science requirements for the two-layer Nautilus small-body program. Values are illustrative design drivers to be refined through end-to-end simulations.\label{tab:scireq}}
\tablewidth{\textwidth}
\tablehead{
\colhead{Requirement} &
\colhead{Layer 1: Cislunar monitoring} &
\colhead{Layer 2: Deep search}
}
\startdata
\reqname{Architecture} &
\reqLone{Candidate distributed wide-field network; illustrative starting point of
$\sim5$--8 $\sim50$\,cm-class nodes in complementary high-Earth and/or cislunar geometries.} &
\reqLtwo{Single NQI unit (a $\sim4$\,m Nautilus Quattro Imaging observatory), scalable to multiple units for greater survey speed and follow-up flexibility} \\
\reqname{Science benchmark} &
\reqLone{Characterize meter- to decameter-scale natural objects traversing the Earth--Moon system through rapid detection and classification, selective
continued tracking, and lunar-impact monitoring. Illustrative design case: sensitivity to $\sim1$\,m potential lunar impactors with hours-scale warning, including the $\sim6$--12\,hr regime relevant to local lunar and LLO applications.} &
\reqLtwo{Recover faint planetary-defense targets and measure the distant population across the Kuiper-Belt-to-Oort-Cloud transition} \\
\reqname{Primary observing mode} &
\reqLone{Wide-field optical imaging in rapid short-frame sequences} &
\reqLtwo{Stable wide-field VIS/NIR imaging for targeted recovery and deep survey fields} \\
\reqname{Bandpass} &
\reqLone{Broad optical discovery band over approximately $0.4$--$1.0\,\mu\mathrm{m}$; color optional} &
\reqLtwo{Broad optical discovery band plus $\geq1$ optical/NIR color; three broad bands preferred for population studies} \\
\reqname{Sensitivity} &
\reqLone{Motion-compensated $m_{\rm lim}\sim21$--22 in broad-survey mode; targeted or longer integrations approaching $m_{\rm lim}\sim23$--24} &
\reqLtwo{Shifted-and-stacked sensitivity reaching $V\sim30$ in the deepest fields} \\
\reqname{Exposure strategy} &
\reqLone{Elementary exposures of order $\sim0.2$--2\,s, combined into $\sim30$--120\,s motion-aware sequences} &
\reqLtwo{Repeated short exposures over several hours or a substantial fraction of a day per discovery field} \\
\reqname{Cadence and revisit} &
\reqLone{Seconds within sequences; minutes--hours for tracklet formation and cross-node association; longer baselines selectively for orbit refinement and continued tracking of high-interest objects} &
\reqLtwo{Days--weeks for confirmation; $\sim3$--4 months for parallax leverage; six months to a year or longer for orbit refinement} \\
\reqname{Field of view and footprint} &
\reqLone{Wide field optimized jointly with apparent-rate coverage, revisit cadence, sensitivity, and number of nodes} &
\reqLtwo{Survey footprint spanning ecliptic and off-ecliptic fields, optimized jointly with depth and revisit cadence} \\
\reqname{Viewing geometry} &
\reqLone{Complementary viewpoints providing strong parallax and resilience to Sun/Earth/Moon background, illumination, and occultation constraints} &
\reqLtwo{Field selection and scheduling designed to exploit changing parallactic displacement} \\
\reqname{Astrometry and dynamics} &
\reqLone{Background-star-tied astrometry, cross-node association, range/orbit estimation, and Earth--Moon three-body trajectory modeling} &
\reqLtwo{Precise relative astrometry with observatory-ephemeris-aware, parallax-aware trajectory modeling} \\
\reqname{Processing and data return} &
\reqLone{Onboard or near-real-time detection, tracklet formation, prioritization, and selective downlink; calibrated elementary exposures retained} &
\reqLtwo{Ground-based synthetic tracking and orbit fitting acceptable; rapid processing and reporting required for planetary-defense targets} \\
\reqname{Rapid response} &
\reqLone{Prioritized trajectory updates and follow-up products for lunar-impact, possible Earth-impact, and other high-interest cislunar objects} &
\reqLtwo{Flexible scheduling for prompt recovery of targets with unresolved impact probabilities} \\
\reqname{Core products and validation} &
\reqLone{Detections, tracklets, trajectory classifications, impact-event products,
and detection/classification/follow-up completeness from injection--recovery and orbit-determination tests} &
\reqLtwo{Astrometry, colors, orbit classifications, and population selection functions from synthetic-tracking and recovery simulations} \\
\enddata
\tablecomments{Layer-1 values define an illustrative design case rather than an optimized architecture or universal operational requirement. Performance
in both layers depends on object brightness and motion, observing geometry, background, detector characteristics, cadence, and processing strategy.
Final requirements should be established through end-to-end geometric-coverage, photometric-yield, injection--recovery, linking, and orbit-determination simulations.}
\end{deluxetable*}

\end{document}